\documentclass[12pt]{iopart}

\newcommand{\rvS}{\Sigma}
\newcommand{\rvXi}{\Xi}

\newcommand{\distS}{p_{\mbox{\tiny $\rvS$}}}
\newcommand{\distXi}{p_{\mbox{\tiny $\rvXi$}}}
\newcommand{\distSXi}{p_{\mbox{\tiny $\rvS \rvXi$}}}

\newcommand{\av}[1]{{\left\langle #1 \right\rangle}}
\newcommand{\fat}[1]{\boldsymbol{#1}}

\usepackage{iopams}
\usepackage{graphicx}

\begin{document}

\title{\bf An optimal Q-state neural network using mutual information}
\author{D Boll\' e and T Verbeiren}
\address{Instituut voor Theoretische Fysica, KU Leuven, B-3001 Leuven, Belgium}
\eads{\mailto{desire.bolle@fys.kuleuven.ac.be}, 
      \mailto{toni.verbeiren@fys.kuleuven.ac.be}}
      
\begin{abstract}
Starting from the mutual information we
present a method in order to find a hamiltonian for a fully connected
neural network model with an arbitrary, finite number of neuron states,
$Q$. For small initial correlations between the neurons and the patterns
it leads to optimal retrieval performance.
For binary neurons, $Q=2$, and biased patterns we recover the
Hopfield model. For three-state neurons, $Q=3$, we find back the recently
introduced Blume-Emery-Griffiths network hamiltonian. We derive its phase
diagram and  compare it with those of related three-state
models. We find that the retrieval region is the largest.
\end{abstract}


\maketitle

One of the challenging problems in the statistical mechanics approach to
associative memory neural networks is the choice of the hamiltonian
and/or learning rule leading to the best retrieval properties including,
e.g., the largest retrieval overlap, loading capacity, basin of attraction,
convergence time.
Recently, it has been shown that the mutual information is the most
appropriate concept to measure the retrieval quality, especially for
sparsely coded networks but also in general (\cite{DB98,BD00} and
references therein).

A natural question is then whether one could use the mutual information in
a systematic way to determine a priori an optimal hamiltonian guaranteeing
the properties described above for an arbitrary scalar valued neuron (spin)
 model.
Optimal means especially that although the network might start initially
{\it far} from the embedded pattern it is  still able to retrieve it.

In the following we answer this question by presenting a general
scheme in order to express the mutual information as a function of the
relevant macroscopic parameters like, e.g., overlap with the embedded
patterns, activity, $\ldots$ and constructing a hamiltonian from it for general
$Q$-state neural networks.
For $Q=2$, we find back the  Hopfield model for biased patterns
\cite{AGS87} ensuring that this hamiltonian is optimal in the sense
described above. For $Q=3$, we obtain a Blume-Emery-Griffiths type hamiltonian
confirming the result found in \cite{DK00}. However, in that paper the
properties of this
hamiltonian have not been discussed, rather the dynamics for an
extremely diluted
version of the model has been treated. Hence, we derive the thermodynamic
phase diagram for the fully connected network modeled by
this hamiltonian and show, e.g., that it has the largest
retrieval region compared with the other three-state models known 
in the literature.

\nosections
Consider a network of $N$ neurons
$\Sigma_i$,  $i=1, \dots, N$, taking different values, $\sigma_i$, from
a discrete set of $Q$ states, ${\cal S}$, with a certain probability 
distribution. In this network we want to
store $p=\alpha N$ patterns $\Xi^{\mu}_i$, $\mu=1, \dots, p$, taking different
values, $\xi^{\mu}_i$, out of the same set ${\cal S}$ with a  certain 
probability distribution. Both sets of random variables are  chosen to
be independent identically distributed with respect to $i$ 

We want to study the mutual information between the neurons and the
patterns, a measure of the correlations between them. At this point we
note that, since the interactions are of infinite range, the
neural network  system is mean-field such that the probability distributions
of all the  neurons and all the patterns are of product type, e.g.,
$p(\{\sigma_i\})=\prod_i p(\sigma_i)$. Furthermore, in a statistical mechanical
treatment  any order parameter $O^\mu$, being a function of
the neurons and the patterns, can be written in the thermodynamic
limit $N \rightarrow \infty$, as
\begin{equation}
O^\mu
 = \lim_{N \to \infty}
      \frac{1}{N} \sum_{i=1}^N O(\sigma_i,\xi_i^\mu)
 = \sum_{\sigma \in {\cal S}} \sum_{\fat{\xi} \in {\cal S}^p} 
            \distSXi(\sigma,\fat{\xi}) 
	                        \, O(\sigma, \xi^\mu)
 \label{form:lln}
\end{equation}
where the left hand side is the configurational average, 
$\fat{\xi}= \{\xi^\mu\}$ and where 
$\distSXi(\sigma,\fat{\xi})$ is the joint
probability distribution of the neurons
and the patterns. Hence, we can forget about the index $i$ in the sequel.

The mutual information $I$ for the random variables $\Sigma$ and $\Xi$ is
then  given by (see, e.g., \cite{Bl90})
\begin{equation}
I(\rvS, \rvXi) =
  \sum_{\sigma\in {\cal S}} \sum_{\fat{\xi}\in {\cal S}^p}
       \distSXi(\sigma,\fat{\xi})
   \ln
      \left(
         \frac{\distSXi(\sigma,\fat{\xi})}
           { \distS(\sigma) \, \distXi(\fat{\xi})}
          \label{def}
  \right) .
\end{equation}

A good network would be one that starts initially {\em far} from a pattern
but is still able to retrieve it. Far in this context means that the
random variables, neurons and patterns, are almost independent.
When $\rvS$ and $\rvXi$ are completely independent, then $\distSXi =
\distS \distXi$. Consequently, when they are almost independent we can
write
\begin{equation}
   \distSXi = \distS \distXi + \Delta_{\mbox{\tiny $\rvS\rvXi$}}
                \label{delta}
\end{equation}
with $\Delta_{\mbox{\tiny $\rvS\rvXi$}}$ small pointwise. We remark that
\begin{equation}
   \sum_{\sigma,\fat{\xi}}
     \Delta_{\mbox{\tiny $\rvS\rvXi$}}(\sigma,\fat{\xi}) = 0 \, .
    \label{delta0}
\end{equation}
Plugging the relation (\ref{delta}) into the
definition (\ref{def}) and expanding the logarithm up to second order
in the small correlations, $\Delta$,  we find using (\ref{delta0})
\begin{equation} \label{form:mutinf:result}
\fl I(\rvS, \rvXi)
= \frac{1}{2} \sum_{\sigma, \fat{\xi}}
     \frac{\left(\Delta_{\mbox{\tiny $\rvS\rvXi$}}(\sigma, \fat{\xi})\right)^2}
          {\distS(\sigma) \distXi(\fat{\xi})}+ \Or(\Delta^3)
  = \frac{1}{2}
      \av{
        \av{
           \left(\Delta_{\mbox{\tiny $\rvS\rvXi$}}(\sigma, \fat{\xi})\right)^2
        }_{\sigma}
      }_{\fat{\xi}}  + \Or(\Delta^3)
\end{equation}
with obvious notation. This approximation is in fact very natural. It is the
average over the square of the difference between the correlated and
uncorrelated probability distribution.

We remark that still all the patterns are contained in
(\ref{form:mutinf:result}). Without loss of generality we consider only
one condensed paterns and omit the index $\mu$ in the sequel.
Consequently, only first and second order correlations of
the variables will be used, higher order ones can be neglected.

Next, we want to express $\Delta_{\mbox{\tiny $\rvS\rvXi$}}$ in terms of
macroscopic, physical quantities of the system (order parameters).
Refering to (\ref{form:lln}) we write down the following $Q^2$
moments
\begin{equation}
m^{cd} = \lim_{N \to \infty} \frac{1}{N} \sum_{i} \sigma_i^c \xi_i^d
       = \av{\sigma^c \, \xi^d}_{\{\sigma, \xi\}}
       = \sum_{\sigma,\xi} \distSXi(\sigma,\xi) \, \sigma^c \, \xi^d
\end{equation}
with $c,d = 0, \dots, Q-1$ and using the notation $0^0=1$. 
We remark that $m^{00}=1$ such that we have
in general $Q^2-1$ independent parameters specifying
$\distSXi(\sigma,\xi)$.

Up to now the derivation is valid for general $Q$-state scalar-valued
neurons. To fix the ideas we choose the neuron states as
\begin{equation}
\sigma_c = -1+\frac{c-1}{Q-1} \qquad \mbox{with }  c = 1, \dots, Q \, .
\end{equation}
This choice corresponds to a Q-state Ising-type architecture leading to
\begin{equation} \label{form:def:T}
\fl
m^{cd} = \sum_{x,y=1}^Q \ T^{cdxy} \ \distSXi(\sigma_x,\xi_y) \ ,
\quad \mbox{with }
T^{cdxy} = \left(
                 -1+\frac{x-1}{Q-1}
              \right)^c
          \left(
             -1+\frac{y-1}{Q-1}
              \right)^d \ .
\end{equation}
In a similar way  we introduce $A$ by
\begin{equation}
\fl
m^{c0}  = \sum_{x} A^{cx} \distS(\sigma_x), \quad
m^{0c}  = \sum_{x} A^{cx} \distXi(\xi_x), \quad
A^{cx} = \left(
           -1+\frac{x-1}{Q-1}
         \right)^c  \, .
\label{form:def:A}
\end{equation}
Finally, by introducing the inverse transformations $S$ and $B$
\begin{equation} \label{form:def:inverse}
\sum_{x,y} T^{cdxy} S_{xyc'd'} = \delta_{a,a'} \, \delta_{b,b'}, \quad
\sum_{x} A^{cx} B_{xd} = \delta_{c,d} \, ,
\end{equation}
we can write the approximation of the mutual information up to order
$\Delta^2$ as
\begin{equation} \label{form:mutinf:orderpar}
\tilde{I}
=
\frac{1}{2}
\sum_{x,y}
\frac
{
\left(
   \sum_{cd} S_{xycd} \, m^{cd}
    - \sum_{cd} \, B_{xc} \, B_{yd} \, m^{c0} \, m^{0d}
\right)^2
}
{
    \sum_{cd} \, B_{xc} \, B_{y
        d} \, m^{c0} \, m^{0d}
}
\end{equation}
where we have left out the dependence on $\rvS$ and $\rvXi$.

Using (\ref{form:lln}), the expression (\ref{form:mutinf:orderpar})
for $\tilde{I}$ can be written in terms of  configurational averages of
the system. In this way, for large $N$, we can express $\tilde{I}$
as a function of the microscopic variables $\sigma_i$ and $\xi_i$. 
Using (\ref{form:mutinf:orderpar}) for every pattern $\mu$, summing over
 $\mu$ and multiplying by $N$ we get an
extensive  quantity, denoted by $\tilde{I}_N$, which grows
monotonically as a function of the correlation between spins and patterns.
Therefore, $H =- \tilde{I}_N$ is a good candidate for a hamiltonian.

Configurational averages
also enter into the denominator of (\ref{form:mutinf:orderpar}). Since we are
mainly interested in the correlations between spins and patterns, rather
than in the respective single probability distributions we assume that the
latter are equal such that we use the known distribution of
the patterns in the denominator.

What we have presented up to now is a scheme to calculate a
hamiltonian  for a general $Q$-state
network with an Ising-type architecture using mutual information.

Next, we discuss some specific examples, $Q=2$ and $Q=3$, in detail.
We start with $Q=2$ states. Given the probabilities associated with each
state, the inversion of  the transformations $T$ (\ref{form:def:T}) and
$A$  (\ref{form:def:A}) leads to
\begin{eqnarray}
\fl
\distSXi(\sigma,\xi) =
   \frac{1-m^{10}-m^{01}+m^{11}}{4} \ \delta_{\sigma,-1} \, \delta_{\xi,-1}
 + \frac{1-m^{10}+m^{01}-m^{11}}{4} \ \delta_{\sigma,-1} \, \delta_{\xi,1}
  \\
\lo
+ \frac{1+m^{10}-m^{01}-m^{11}}{4} \ \delta_{\sigma,1} \, \delta_{\xi,-1}
 + \frac{1+m^{10}+m^{01}+m^{11}}{4} \ \delta_{\sigma,1} \, \delta_{\xi,1}
  \, .
  \nonumber
\end{eqnarray}
The distributions $\distS(\sigma)$ and $\distXi(\xi)$ can be found by
summing out $\xi$ and  $\sigma$ respectively. Using these distributions,
 $\tilde{I}$ becomes
\begin{equation}
\tilde{I}
= \frac{\left( m^{11}-m^{01}m^{10}
                    \right)^2}{2(1-(m^{01})^2)(1-(m^{10})^2)}\, .
\end{equation}
Substituting the averages over the probability distributions by
configurational averages and putting $m^{10}= m^{01}=b$ in the
denominator, where b is the bias of the patterns, we get 
\begin{equation}
H = -\frac{1}{2(1-b^2)^2} N \sum_\mu
        \left(
          \frac{1}{N} \sum_{i} \sigma_i \xi_i^\mu - b \sigma_i
        \right)^2 \, .
\end{equation}
This hamiltonian can be written as
\begin{equation}
\fl
H= -\frac{1}{2} \sum_{ij} J_{ij} \sigma_i \sigma_j
       \quad \mbox{with} \quad
    J_{ij}
        = \frac{1}{N(1-b)^2} \sum_{\mu} (\xi_i^\mu-b)(\xi_j^\mu-b)
\end{equation}
and this is precisely the Hopfield hamiltonian with \cite{AGS87} and
without \cite{Hop82} bias.

We remark that a particularly nice aspect of this treatment is that the
adjustment of the learning rule due to the bias enters in a
natural way.
Furthermore, we learn that the Hopfield hamiltonian is the optimal
two-state hamiltonian in the sense that we started from the
mutual information calculated for an initial state having a small overlap
with the condensed pattern. This confirms a well-known fact in the
literature.

One could ask what happens when one assumes that initially the
state of the network is already {\it close} to the embedded pattern. Since
the mutual information for fully correlated random
variables is equal to the entropy, $S(\rvXi), $\cite{Bl90} one is 
interested in (assuming again one condensed pattern)
$F(\rvS,\rvXi) = I(\rvS,\rvXi) - S(\rvXi) \, $.
We define
\begin{equation}
\fl
\distSXi(\sigma, \xi) = \sum_{\sigma', \xi'} \left[
       \distSXi^d(\sigma', \xi') \,
             \delta_{\sigma',\sigma}\delta_{\xi',\xi}
       + \distSXi^{od}(\sigma', \xi') 
        (1-\delta_{\sigma',\sigma}\delta_{\xi',\xi}) \right]
\end{equation}
with obvious notation. Writing
\begin{equation}
\distSXi^d = \distS \, + \Delta'_{\mbox{\tiny $\rvS \rvXi$}}
\end{equation}
with $\Delta'_{\mbox{\tiny $\rvS \rvXi$}}$ small pointwise for large 
correlations and assuming that
$\distSXi^{od}(\sigma, \xi)=0$ for $\forall \sigma, \xi$, in order to
retain only the polynomial behaviour,  we expand $F$ and find 
\begin{equation} \label{form:mutinf:lc}
F(\rvS,\rvXi) = \frac{1}{2} \sum_{\sigma,\xi} 
        \frac{(\Delta'_{\mbox{\tiny $\rvS \rvXi$}}(\sigma,\xi))^2}
              {\distS(\sigma)} 
                      + {\Or}({\Delta'}^3) \, .
\end{equation}
Expressing $F$ in terms of the order parameters as in 
(\ref{form:mutinf:orderpar}), we get the hamiltonian \cite{AI90,BHS94}
\begin{equation}
  H= N(1-b^2)\prod_\mu (1-m_\mu^2) \quad 
           \mbox{with} \quad m_\mu= \frac{1}{N} \sum_i 
         \frac{(\xi_i^\mu -b) \sigma_i}{1-b^2}
\end{equation}
for one pattern.  In  \cite{BHS94} it is shown that this
hamiltonian can store an infinite number of patterns.  This is consistent with
the intuitive idea that it is possible to store a lot of patterns as long
as the network state is initially close to them.

For $Q=3$ we focus, without loss of generality, on the case where the
distributions are taken symmetric around zero, meaning that all the odd
moments vanish.  Following the scheme proposed above we arrive at 
\begin{equation}
\fl \tilde{I} = \frac{1}{2} \, \frac{1}{m^{02}m^{20}} \, (m^{11})^2
    + \frac{1}{2} \, \frac{1}{m^{02} m^{20} (1-m^{02}) (1-m^{20})}
           (m^{22} - m^{02}m^{20})^2 \, .
\end{equation}
Identifying $m^{02}=a$ as the activity of the patterns, $m^{20}=q$ as
the activity of the neurons, $m^{11}=m$ as the overlap, $m^{22} = n$ as
the activity overlap \cite{BD00}, and defining $l = n-aq$ we arrive at
\begin{equation}
\tilde{I} = \frac{1}{2} \, \frac{1}{a^2} \, m^2
    + \frac{1}{2} \, \frac{1}{(a(1-a))^2} \, l^2
    \ .
\end{equation}
This leads to a hamiltonian
\begin{equation}                       \label{form:hamiltonian}
H = - \frac{1}{2} \sum_{i, j} J_{ij} \sigma_i \sigma_j
    - \frac{1}{2} \sum_{i, j} K_{ij} \sigma_i^2 \sigma_j^2
\end{equation}
with
\begin{equation}
\fl  J_{ij}  = \frac{1}{a^2N} \sum_{\mu=1}^p \xi^\mu_i \xi^\mu_j \, , \quad
  K_{ij}  = \frac{1}{(a(1-a))^2N} \sum_{\mu=1}^p \eta^\mu_i \eta^\mu_j
  \, , \quad \eta^\mu_i= (\xi^\mu_i)^2 - a \ .
\end{equation}
This hamiltonian resembles the Blume-Emery-Griffiths (BEG)
hamiltonian \cite{BEG71}. The derivation above confirms the result found
in \cite{DK00} starting from an explicit form of the mutual information
for $Q=3$.
In that paper the dynamics has been studied for an extremely
diluted asymmetric version of this model.
Here we want to discuss the fully connected
architecture and derive the thermodynamic phase diagram, which has not
been done in the literature, in order to compare it with the other $Q=3$
state models known.

In order to calculate the free energy we use the standard replica
method \cite{MPV78}. Starting from the replicated partition function
and assuming replica symmetry we obtain
\begin{eqnarray}
\fl
f  =  \frac{1}{2} \sum_\nu 
            \left( m_\nu^2
                   + l_\nu^2
            \right)
+ \frac{\alpha}{2\beta} \log (1-\chi)
+ \frac{\alpha}{2\beta} \log (1-\phi)
+ \frac{\alpha}{2\beta} \frac{\chi}{1-\chi} \nonumber \\
\fl \qquad + \frac{\alpha}{2\beta} \frac{\phi}{1-\phi}
+ \frac{\alpha}{2} \frac{A q_1 \chi}{(1-\chi)^2}
+ \frac{\alpha}{2} \frac{B p_1 \phi}{(1-\phi)^2} 
- \frac{1}{\beta}
\left\langle \int Ds Dt \log \Tr_\sigma \exp \left ( \beta \tilde H \right )
\right\rangle_{\{\xi^\nu\}} 
\end{eqnarray}
with $\nu$ denoting the condensed patterns and
\begin{equation} 
\fl
\tilde H  = A \sigma
 \left[
  \sum_\nu m_\nu \xi^\nu
     + \sqrt{\alpha r}s
 \right]
   + B \sigma^2
 \left[
\sum_\nu l_\nu \eta^\nu
+  \sqrt{\alpha u} t
\right]
+ 
 \frac{\sigma^2 \alpha A \, \chi}{2(1-\chi)} 
+ 
 \frac{\sigma^4 \alpha B \, \phi}{2(1-\phi)} 
\end{equation}
and $A=1/a$, $B=1/(a(1-a))$, $Ds$ and $Dt$ Gaussian measures, $Ds=ds
(2\pi)^{-1/2} \exp(-s^2/2)$, and
\begin{equation}
\fl
\chi = A \beta ( q_0 - q_1 )\, , \quad  \phi = B \beta ( p_0 - p_1)\, ,
   \quad r = \frac{q_1}{(1-\chi)^2}\, ,
   \quad u = \frac{p_1}{(1-\phi)^2} \, .
\end{equation}
For $Q=3$ ($\sigma^2=\sigma^4$), the order parameters are defined as follows
\begin{equation}
\fl
\begin{array}{rclrcl}
m_\nu & = &
A \left\langle  \xi^\nu \int \!  Ds Dt \  
\left\langle \sigma \right\rangle_\beta
\right\rangle_{\{\xi^\nu\}}
&\qquad
q_1 & = &
\left\langle
\int \! Ds Dt \  {\left\langle \sigma
\right\rangle}_\beta^2
\right\rangle_{\{\xi^\nu\}}
\\
l_\nu & = &
B\left\langle \eta^\nu \int \! Ds Dt \  \left\langle \sigma^2
\right\rangle_\beta
\right\rangle_{\{\xi^\nu\}}
&\qquad 
p_1 & = &
\left\langle
\int \! Ds Dt \  {\left\langle
\sigma^2 \right\rangle}^2_\beta
\right\rangle_{\{\xi^\nu\}}
\\
q_0 &  = &p_0 =
\left\langle
\int \! Ds Dt \  \left\langle \sigma^2 \right\rangle_\beta
\right\rangle_{\{\xi^\nu\}}
\end{array}
\end{equation}
with the small brackets $\langle \ldots \rangle_\beta$ denoting the
usual thermal average.
We recall that $m_\nu$ is the overlap, $l_\nu$
is related to the activity overlap, $q_0$ is the activity of the neurons and
$q_1$ and $p_1$ are Edwards-Anderson parameters. For one condensed pattern
the index $\nu$ can be dropped.
\noindent
\begin{figure}[tb]
 \centerline{
 \includegraphics[width=.5\textwidth,clip=]{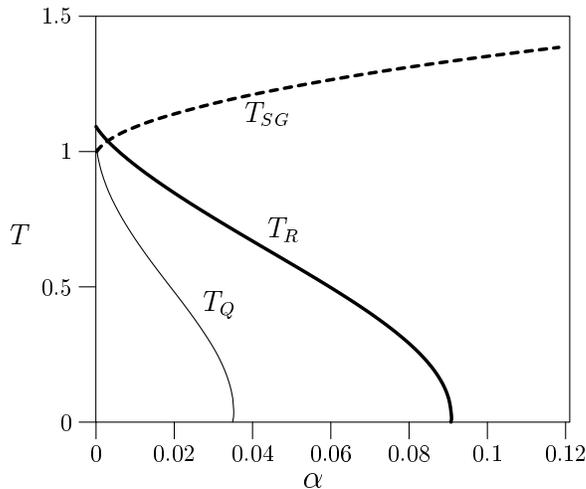}
 }
 \begin{center}
 \caption{$Q=3$ $T-\alpha$ phase diagram for uniform patterns.  The
 meaning of the lines is explained in the text.}
 \label{pd:a2__3}
 \end{center}
\end{figure}

Solving the fixed-point equations for the order parameters and considering
uniform patterns ($a=2/3$), we obtain a rich $T-\alpha$ phase diagram
(see \cite{BV} for more details). The phases that are important from a
neural network point of view are presented in  \fref{pd:a2__3}.
The border of the retrieval phase ($m>0$, $l>0$) is denoted by a thick full
line.  The most
important result is that the capacity of the BEG neural network is much
larger than that of other $Q=3$ models. Compared with the
$Q=3$-Ising model \cite{BRS94}, e.g., it is almost twice at $T=0$.  
 Of course this is due
to the second term in the  hamiltonian (\ref{form:hamiltonian}). A
study of the dynamics of this model, which is in progress, confirms
this result.
Another new feature in the phase diagram, compared with other models,
is the so-called quadrupolar phase  ($m=0$ but $l>0$) which lies below  
the thin full line. It is present in the original BEG spin-model
\cite{BEG71} and has also been seen for the extremely diluted network
model \cite{DK00}.
In this phase the active neurons ($\pm 1$) coincide with the active
patterns but the sign does not. This means that although the system does
not succeed in  retrieval the information content is nonzero.
For $a=2/3$ this phase lies completely within the retrieval phase
but for other values of $a$ (e.g., $a=0.8$) it does not \cite{BV}.
Besides these phases one also has a spin-glass phase and a
paramagnetic phase (separated by the broken line in
\fref{pd:a2__3}). The latter coexists with the retrieval phase in a
region near the $T$-axis. We refer to \cite{BV} for further details.

\nosections
In conclusion, we have presented a method starting from the mutual
information beween the neurons and the patterns to derive an optimal
hamiltonian for a general $Q$-state neural network.
The derivation assumes that the correlations
between the neurons and patterns are small initially, and thus guarantees
optimal retrieval properties (loading capacity, basin
of attraction) for the model.  For $Q=2$, we find back the
Hopfield hamiltonian for biased patterns, while for $Q=3$ we find
the Blume-Emery-Griffiths hamiltonian.
We have derived the phase diagram for this fully connected BEG
model confirming that the capacity is larger than the one for
related models. We believe that similar results can be obtained for
vector models
and other architectures. An extended version of the work on the BEG
fully connected neural network will appear in~\cite{BV}.

\ack

We would like to thank D. Dominguez and G.M. Shim for useful discussions.
This work was supported in part by the Fund for Scientific Research,
Flanders (Belgium).

\end{document}